\documentclass[a4paper,twocolumn,english,prl,superscriptaddress]{revtex4}
\usepackage{geometry}
\usepackage{array}
\usepackage{amsmath}
\usepackage{amsfonts}
\usepackage{graphicx}
\usepackage{xcolor}
\usepackage{bm}
\usepackage{physics}
\usepackage{hyperref}
\usepackage[all]{hypcap}

\hypersetup{
	colorlinks=true,
	linkcolor=blue,
	filecolor=gray,      
	urlcolor=blue,
	citecolor=blue,
}

\begin{document}
\title{Dynamic generation of spin spirals of moir\'e trapped carriers via exciton mediated spin interactions}
\author{Chengxin Xiao}
\affiliation{Department of Physics, The University of Hong Kong, Hong Kong, China}
\affiliation{HKU-UCAS Joint Institute of Theoretical and Computational Physics at Hong Kong, China}
\author{Yong Wang}
\affiliation{School of Physics, Nankai University, Tianjin, China}
\affiliation{Department of Physics, The University of Hong Kong, Hong Kong, China}
\author{Wang Yao}
\email{wangyao@hku.hk}
\affiliation{Department of Physics, The University of Hong Kong, Hong Kong, China}
\affiliation{HKU-UCAS Joint Institute of Theoretical and Computational Physics at Hong Kong, China}
\date{\today}
\begin{abstract}
    Stacking transition metal dichalcogenides (TMDs) to form moir\'e superlattices has provided exciting opportunities to explore many-body correlation phenomena of the moir\'e trapped carriers. TMDs bilayers, on the other hand, host long-lived interlayer exciton (IX), an elementary excitation of long spin-valley lifetime that can be optically or electrically injected. Here we find that, through the Coulomb exchange between mobile IXs and carriers, the IX bath can mediate both Heisenberg and Dzyaloshinskii-Moriya type spin interactions between moir\'e trapped carriers, controllable by exciton density and exciton spin current respectively. We show the strong Heisenberg interaction, and the extraordinarily long-ranged Dzyaloshinskii-Moriya interaction here can jointly establish robust spin spiral magnetic orders in Mott-Wigner crystal states at various filling factors, with spiral direction controlled by exciton current.     
\end{abstract}
\maketitle

\textit{Introduction} -- Stacking transition metal dichalcogenides (TMDs) monolayers to form moir\'e superlattice has provided an arena to explore many-body correlation phenomena of the spin-valley locked massive Dirac fermions~\cite{Regan2020,Xu2020,Tang2020,Huang2021,Miao2021,PhysRevLett.127.037402,Shimazaki2020,Zhou2021,Ghiotto2021,Li2021}.
Experiments have discovered correlated insulating states at various integer and fractional fillings~\cite{Regan2020,Xu2020,Tang2020,Huang2021,Miao2021,PhysRevLett.127.037402,Shimazaki2020,Zhou2021}, attributed to charge ordering by strong Coulomb interaction in the moir\'e energy landscape, namely Mott and generalized Wigner crystal states.
Their magnetic properties have also been explored using magneto-optical measurements in WSe$_2$/WS$_2$ heterobilayers~\cite{Tang2020,Wang2022}. Super-paramagnetic response is observed at integer hole filling~\cite{Tang2020,Wang2022}, attributed to antiferromagnetic coupling between nearest-neighbour moir\'e sites~\cite{Tang2020}.
Remarkably, reflective magnetic circular dichroism (RMCD) measurement shows evidences of spontaneous ferromagnetic order at fractional hole filling when and only when the optical excitation power exceeds a modest threshold~\cite{Wang2022}, implying spin interaction enabled by the optical excitation~\cite{PhysRevLett.89.167402,PhysRevLett.93.127201}.

Another topic of intensive interest in TMDs heterostructures is the long-living interlayer excitons (IXs) with electron and hole constituents separated to adjacent layers~\cite{Rivera2015, Rivera2016, Unuchek2018, Seyler2019, Tran2019, science.aaw4194, PhysRevLett.123.247402, Brotons-Gisbert2020, Li2020, Bai2020,Tan2021}.
In the type-II band alignment, optical excitation can establish large population and spin-valley polarization of these low energy excitons, utilizing valley selection rule in individual layer~\cite{PhysRevLett.108.196802}, followed by ultrafast interlayer charge transfer~\cite{Ceballos2014, Lee2014}.
And ultralong lifetime and spin-valley lifetime of IXs are observed~\cite{Rivera2016,Tan2021}.
Like the charge carriers, IXs can also experience a periodic potential from the moir\'e landscape.
Experiments in MoSe$_2$/WSe$_2$ moir\'e show that at very low excitation power and low temperature, IX can get trapped by the moir\'e potential, exhibiting quantum dot like narrow resonances in photoluminescence~\cite{Seyler2019, Brotons-Gisbert2020, Li2020}.
At modest excitation or higher temperature (30K), the narrow-line emission is quenched and dominated by broad peaks, implying that traps are rather shallow and IXs become mobile under these conditions~\cite{Seyler2019, Bai2020}. The ability to drive IX flow by electrical control has been demonstrated~\cite{Unuchek2018,sciadv.aba1830}.
Moreover, condensation of electrically injected IXs~\cite{Wang2019}, and equilibrium IXs in excitonic insulating states~\cite{Chen2022, Zhang2022} have also been reported in TMDs heterostructures.

Here we discover a novel form of spin-spin interaction between moir\'e trapped carriers through the mediation by a bath of mobile IXs in bilayer TMDs.
The exchange between trapped electron (hole) and the electron (hole) constituent of IX forms the basis of a Ruderman-Kittel-Kasuya-Yosida (RKKY) type interaction between trapped carriers through the mediation by mobile IXs.
With an IX bath in Bose distribution, the mediated interaction between trapped carriers has a Heisenberg form with interaction range and strength controlled by density and temperature of IXs, capable of establishing ferromagnetic order at carrier-carrier distance up to $10-20$ nm as observed in Ref.~\cite{Wang2022}.
Non-equilibrium part of IX distribution that corresponds to an excitonic spin current can further introduce Dzyaloshinskii-Moriya interaction (DMI) of extraordinarily long range, practically limited by exciton's coherence length only. By this long-range and anisotropic DMI, albeit weak, the ferromagnetic order gets pinned in-plane and spirals along the excitonic current direction with sub-micron wavelength, where the critical temperature of forming the magnetic spiral is determined by the strong Heisenberg part. Our finding points to an exciting possibility to dynamically establish and manipulate magnetization textures in Mott-Wigner crystal states in moir\'e.


\begin{figure}[htpb]
    \centering
    \includegraphics[width=8cm]{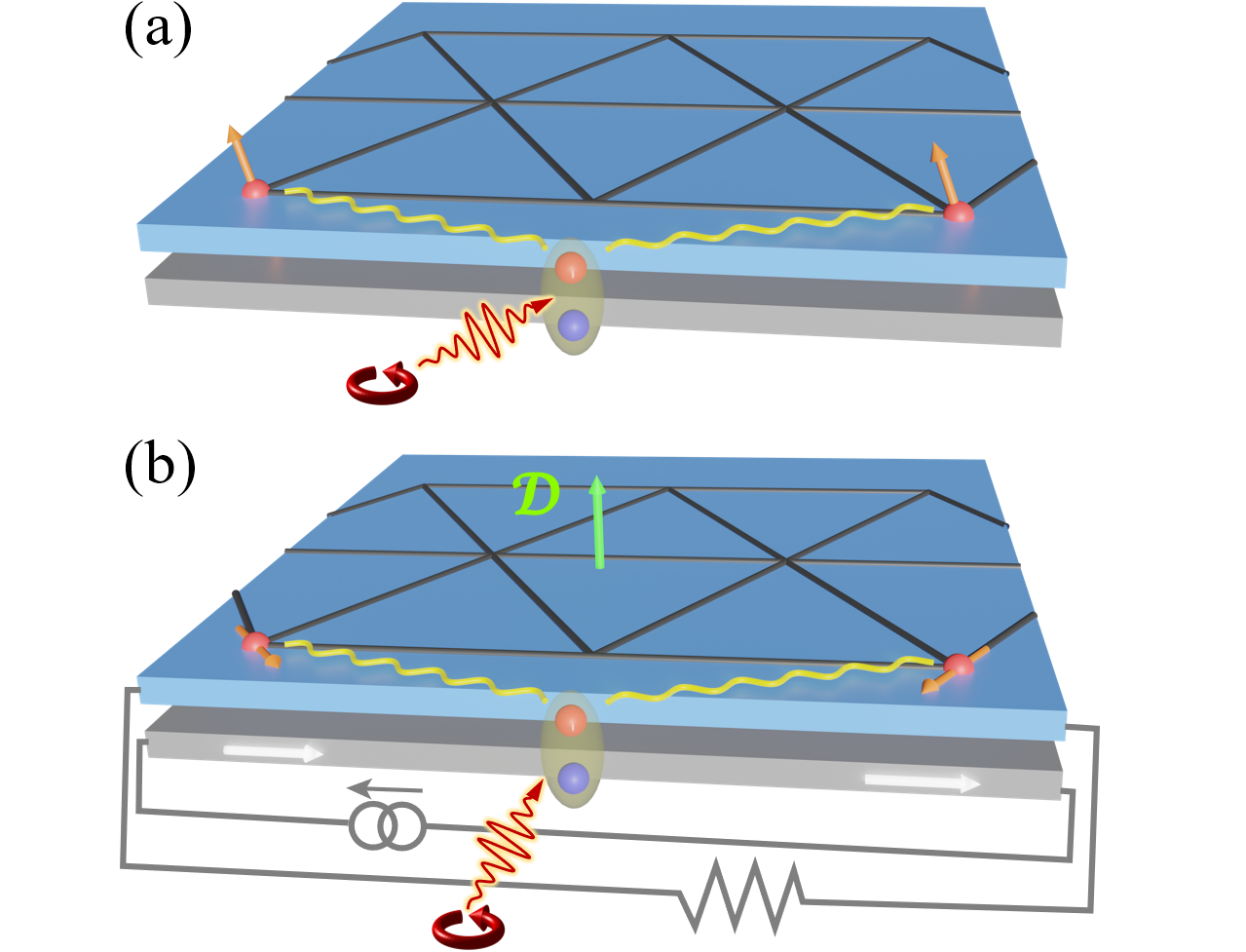}
    \caption{(a) Through the mediation by mobile interlayer excitons (IX), a Heisenberg interaction $\mathcal J \bm{S}_{i} \cdot \bm{S}_{j} $ is realized for moir\'e  trapped holes (red spheres on triangular lattice sites). Wavy lines denote Coulomb exchange between trapped holes and the hole constituent of IX. (b) In presence of an IX spin current, a long-range Dzyaloshinskii-Moriya interaction $\bm{\mathcal D} \cdot (\bm{S}_{i} \times \bm{S}_{j}) $ is also introduced between the trapped holes, with an out-of-plane $\bm{\mathcal D}$. The schematic illustrates optical injection of spin polarized IX whose lateral motion is driven by electrical current.
    }\label{fig1}
\end{figure}

We consider a TMD heterobilayer doped with holes, which are trapped in the triangular moir\'e superlattice potential formed in the top layer due to the twisting and/or lattice mismatch (Fig.~\ref{fig1}). A bath of spin polarized mobile IXs coexist in the system either by optical or electrical injection. The trapped hole can exchange with the hole constituent of the IX residing in their common layer, leading to a hole-exciton interaction dependent on their spin configurations. The Hamiltonian for the trapped holes and mobile IXs can be generally written as,
\begin{eqnarray}
    H_0&=& \sum_{\bm{k} \sigma_e\sigma} \epsilon_{\bm{k}\sigma_e\sigma} a_{\bm{k}\sigma_e\sigma}^{\dagger} a_{\bm{k}\sigma_e\sigma}+\sum_{j \sigma} \epsilon_{d} d_{j \sigma} ^{\dagger} d_{j \sigma},  \notag                                   \\
    H_v&=&\sum_{\begin{subarray}{c}
            j \bm{k} \bm{k}^{\prime}
            \sigma \sigma^{\prime}\sigma_e
        \end{subarray}}I e^{-i(\bm{k}-\bm{k}^{\prime})\cdot\bm{R}_j} d^{\dagger}_{j\sigma^{\prime}} a_{\bm{k}  \sigma_e\sigma}^{\dagger}a_{\bm{k}^{\prime} \sigma_e \sigma^{\prime}} d_{j\sigma}. \notag
\end{eqnarray}
$a^{\dagger}_{\bm{k}\sigma_e\sigma}$ creates an IX with center-of-mass momentum $\bm{k}$ and spin indices $\sigma_{e}$ and $\sigma$ for its electron and hole constituents respectively, and $d^\dagger_{j\sigma}$ creates a trapped hole of spin index $\sigma$ at lattice site $j$.
$H_v$ describes the Coulomb exchange between the IXs and trapped holes. The magnitude and sign of $I$ are determined by the competition of three scattering channels~\cite{COMBESCOT2008215, PhysRevB.86.115210}, depending on the wavefunction overlap of IX and trapped hole (details in Supplementary~\cite{supplementary}).

\begin{figure*}[htpb]
    \centering
    \includegraphics[width=17cm]{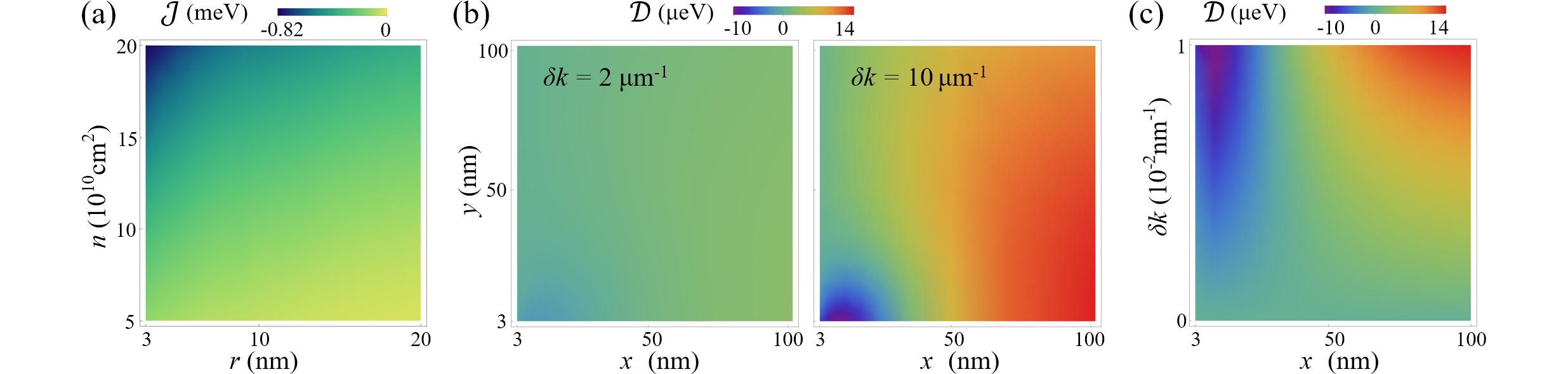}
    \caption{Spin interactions controlled by density (\(n\)) and spin current of the exciton bath.
    (a) The isotropic Heisenberg interaction strength \(\mathcal{J}(\bm{r})\), with the exciton bath in Bose distribution $f_0^{n,T}$ of various density \(n\) at $T=30$K. (b, c) The Dzyaloshinskii-Moriya interaction set on by an exciton spin current, the latter described by a shifted distribution \(f({\bm{k}})=f_0^{n,T}({\bm{k}}+\delta\bm{k})\), with $n=10^{11}$cm$^{-2}$ and $T=30$K. In (b), the anisotropic \(\mathcal{D} (\bm{r})\) is shown for a shift in the $x$ direction by \(\delta k=2\mu\)m\({}^{-1}\) and \(\delta k=10 \mu\)m\({}^{-1}\) respectively. In (c),
    \(\mathcal{D}(x, y=0)\), the interaction along the current direction, is plotted as a function of $\delta k$.} \label{fig2}
\end{figure*}

\textit{Exciton mediated spin-spin Interaction} --
We follow the standard approach to derive the indirect RKKY type interaction between the trapped holes mediated by the mobile IXs. With the interaction $H_{v}$ treated as perturbation, the effective spin-spin interaction between the trapped holes can be obtained by performing the Schrieffer-Wolff transformation $e^{\mathcal{S}}He^{-\mathcal{S}}$, where $\mathcal{S}$ satisfies $[H_{0},\mathcal{S}]=H_{v}$. Tracing out the IX degrees of freedom and keeping to the second order of $H_{v}$, we obtain the effective Hamiltonian for trapped holes,
\begin{equation}
    H_{h} =\sum_{i,j}\mathcal{J}(\bm{r}_{ij}) \bm{S}_{i} \cdot \bm{S}_{j} + \eta \bm{\mathcal{D}}(\bm{r}_{ij}) \cdot (\bm{S}_{i} \times \bm{S}_{j}). \label{HSS}
\end{equation}
The two terms are of Heisenberg and DMI form respectively~\cite{YU20211}, with coupling coefficients,
\begin{eqnarray}
    \mathcal{J}(\bm{r}_{ij}) & =&\sum_{\bm{k} \bm{k}'}\dfrac{I^2}{2}\cos\left[(\bm{k}'-\bm{k})\cdot\bm{r}_{ij}\right]\dfrac{f({\bm{k}})-f({\bm{k}'})}{\epsilon({\bm{k}})-\epsilon({\bm{k}'})}, \notag \\
    \bm{\mathcal{D}}(\bm{r}_{ij}) & =& \hat{\bm z} \sum_{\bm{k} \bm{k}'}\dfrac{I^2}{2}\sin\left[(\bm{k}'-\bm{k})\cdot\bm{r}_{ij}\right]\dfrac{f({\bm{k}})+f({\bm{k}'})}{\epsilon({\bm{k}})-\epsilon({\bm{k}'})}. \notag \\
    \label{DMI}
\end{eqnarray}
Here \(S_{j}^{+}=d^{\dagger}_{j\uparrow}d_{j\downarrow}\), \(S_{j}^{-}=d^{\dagger}_{j\downarrow}d_{j\uparrow}\),
\(S_{j}^{z}=\frac{1}{2}(d^{\dagger}_{j\uparrow}d_{j\uparrow}-d^{\dagger}_{j\downarrow}d_{j\downarrow})\) are the spin operators of the trapped holes. \(\bm{r}_{ij} \equiv \bm{R}_{j}-\bm{R}_{i}\) is the displacement vector from site $i$ to $j$.
Without losing generality, we consider a bath of spin singlet IX in distribution $f_{\uparrow }=\frac{1+\eta}{2}f({\bm{k}})$, $f_{\downarrow}=\frac{1-\eta}{2}f({\bm{k}})$, $\eta$ denoting the exciton spin polarization. The exchange with a trapped hole of a different spin-valley index can turn the singlet to the intervalley spin triplet. With electron-hole exchange quenched by their layer separation, the energy splitting between these two species is negligible. The spin index can therefore be dropped in the exciton dispersion \(\epsilon_{\bm{k}\sigma_e \sigma} = \epsilon ({\bm{k}}) =\frac{\hbar^2 k^2}{2 m_{X}}\).

With the isotropic IX dispersion, it is obvious that the DMI coefficient \(\mathcal{D}\) will vanish if the distribution is an even function \(f({\bm{k}})=f({-\bm{k}})\), as in a thermal distribution.
In this case, there only exists a Heisenberg-type spin-spin interaction in Eq.~(\ref{HSS}), as schematically shown Fig.~\hyperref[fig1]{1(a)}, which favors ferromagnetic order of the trapped holes.
The DMI-type term in~\eqref{HSS} can emerge for a non-equilibrium IX bath with \(\eta\neq 0\) and \(f({\bm{k}})\neq f(-{\bm{k}})\), which corresponds to a spin-polarized exciton current. Flipping the spin polarization and the current direction simultaneously will not change \(\mathcal{D}\), meaning that it can also be induced by a pure spin current of exciton without net spin polarization. In contrast to the DMI interaction of magnetic impurities mediated by equilibrium electrons in spin-momentum locked bands~\cite{PhysRevLett.106.097201, PhysRevB.92.224435}, we note the DMI vector here is out-of-plane \(\bm{\mathcal{D}}=\hat{\bm z}\mathcal{D}(\bm{r}_{ij})\), so this DMI tends to align spins in-plane in orthogonal directions.


Fig.~\ref{fig2} presents examples of the numerically calculated Heisenberg and DMI coefficients \(\mathcal{J}(\bm{r})\) and \(\mathcal{D}(\bm{r})\).
We have set the distribution function as \(f({\bm{k}})=f_0^{n,T}({\bm{k}}+\delta\bm{k})\), where $f_0^{n,T}$ denotes a Bose distribution at temperature $T$ and density $n$,
and the momentum space shift $\delta\bm{k}$ is linearly proportional to the IX current or drift velocity in the range concerned.
The isotropic Heisenberg coefficient \(\mathcal{J}\) as a function of IX density \(n\) and hole-hole separation is shown in Fig.~\hyperref[fig2]{2(a)}, at $T=30$K. In contrast to the RKKY mediated by fermions, the exciton mediated \(\mathcal{J}\) is always negative, i.e.~ferromagnetic coupling, in the concerned parameter range, due to the nature of the Bose distribution. A ferromagnetic order is thus favored for the trapped holes.
Different from the oscillating feature in fermion-mediated RKKY interaction, the magnitude of \(\mathcal{J}\) decreases monotonically with increased distance.
The exchange strength can reach $\sim0.1$meV at a hole-hole separation of 5~nm, under IX density of $10^{11}$cm$^{-2}$, consistent with the observed $T_c$ of the ferromagnetic order and the estimated IX density as reported in Ref.~\cite{Wang2022}. We note that \(\mathcal{J}\) has no significant change for the range of \(\delta k\) considered.

Fig.~\hyperref[fig2]{2(b)} plots \(\mathcal{D} (\bm{r}) \) at $\delta k=2\mu$m$^{-1}$ and $10\mu$m$^{-1}$ respectively, the former corresponding to a drift velocity of 600 m/s (observed for monolayer exciton driven by dynamical strain~\cite{Datta2022}). The DMI coefficient is anisotropic, which oscillates along the current direction (\(x\)) and decay in \(y\) direction. Remarkably, the DMI is extraordinarily long range in the current direction, without noticeable decay in the range shown. This is a characteristic from the Bose distribution in the mediation channel, in sharp contrast to the electron mediated RKKY~\cite{PhysRevLett.106.097201, PhysRevB.92.224435}. Fig.~\hyperref[fig2]{2(c)} shows \(\mathcal{D}\) along the current direction, at various \(\delta k\).


\(\mathcal{D} (\bm{r}) \) calculated using Eq.~(\ref{DMI}) has assumed that IX retains its phase coherence between successive scatterings with trapped holes. With a finite phase coherence length $l_{\phi}$ in reality,  \(\mathcal{D} (\bm{r}) \) shall be multiplied with a factor $e^{-r/l_{\phi}}$, which determines its practical interaction range along the current direction. $l_{\phi}$ of hundreds of nm has been measured for quantum well excitons~\cite{PhysRevLett.89.097401}.
This suggests that even with significant decoherence, the DMI interaction range is still much larger than the Heisenberg. While the peak value of \(\mathcal{D}\) is  nearly two orders of magnitude smaller than that of \(\mathcal{J}\), the anisotropic DMI here can nevertheless have considerable contribution to the total energy in determining the magnetic order.

\begin{figure*}[htpb]
    \centering
    \includegraphics[width=17cm]{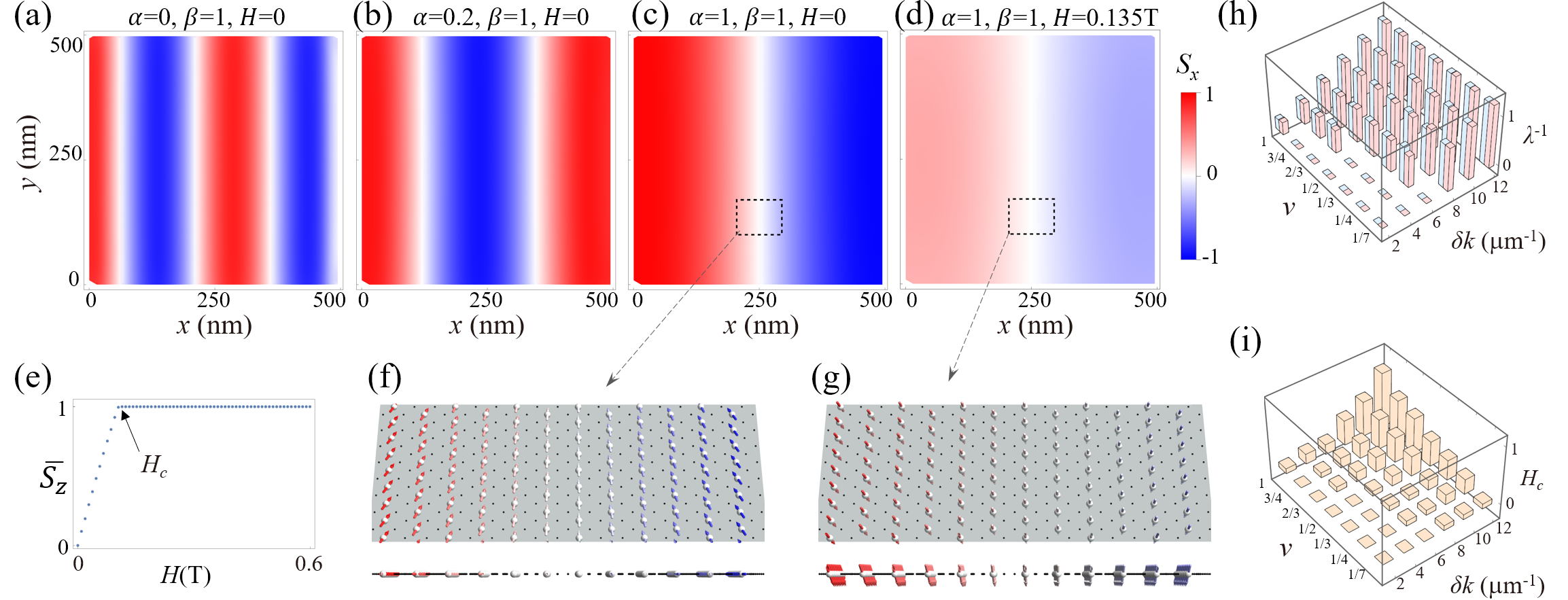}
    \caption{(a-g) Spin textures of an ordered \( \nu=1/3\) hole filling, from the DMI mediated by an exciton bath in distribution \(f({\bm{k}})=f_0({\bm{k}}+\delta\bm{k})\), with the shift \(\delta k=10\mu\)m\({}^{-1}\). $f_0$ is a Bose distribution of $n=10^{11}$cm$^{-2}$ and $T=30$K.
    \(x\)-component of the spins are color coded. The Heisenberg interaction is gradually switched on from zero in (a) to full strength in (c) and (d), as denoted by parameter \(\alpha\) (c.f. Eq.~(\ref{hamil})). (d) has a finite magnetic field $H=0.135$T applied out-of-plane.
    (e) shows the out-of-plane magnetization curve, and $H_c$ denotes the critical field of its saturation. (f) and (g) show zoom-in plots of the three-dimensional spin configurations within dashed box in (c) and (d) respectively. (h) The spiral wavevector \(\lambda^{-1}\) in units of \(\mu m^{-1}\), and (i) the critical field \(H_c\) in units of Tesla, as functions of $\delta k$ and filling factor \(\nu\) of the charge order. In (h), red and blue bars are for exciton current in the $x$ and $y$ directions respectively.}\label{fig3}
\end{figure*}

\textit{Current controlled magnetic spirals} -- Conventional DMI tends to induce non-collinear magnetization textures such as spin spirals or skyrmions in chiral magnetic materials~\cite{YU20211}. In the Mott-Wigner crystal states of the trapped holes, we show that the two distinct parts of IX mediated interaction together establishes robust long wavelength magnetic spirals along the current direction.

In search of the ground state spin configuration, we exploited the classical Monte Carlo simulation, performed on a $500\times500$~nm triangular lattice with open boundary condition, where the moir\'e superlattice constant is $5$~nm (see Supplementary~\cite{supplementary}).
The decoherence of IX is taken into account by multiplying $\mathcal{J}$ and $\mathcal{D}$ by the factor $e^{-r/l_{\phi}}$ with coherence length $l_{\phi} = 30$~nm.
A perpendicular magnetic anisotropy of $3\mu$eV is extracted from the coercive field measured in Ref.~\cite{Wang2022}.
In examining the magnetic field effect, an out-of-plane g-factor of 10 is taken for the holes~\cite{Aivazian2015}.
The numerical simulations are performed with the following Hamiltonian,
\begin{equation}
    H_{h}' =\sum_{i,j}\alpha\mathcal{J}(\bm{r}_{ij}) \bm{S}_{i} \cdot \bm{S}_{j} +\beta \eta {\mathcal{D}}(\bm{r}_{ij}) (\bm{S}_{i} \times \bm{S}_{j})_{z}. \label{hamil}
\end{equation}
where the parameters \(\alpha\) and \(\beta\) are  introduced to artificially switch on/off the two terms for examining their individual roles at given IX bath parameters.

Fig.~\hyperref[fig3]{3(a-c)} shows the ground state magnetization texture at hole filling $\nu = 1/3$, where the IX bath has a density $n=10^{11}$~cm$^{-2}$
and $\delta k=10\mu$m$^{-1}$. The Heisenberg interaction is progressively switched on, from zero in Fig.~\hyperref[fig3]{3(a)}, to full strength in \hyperref[fig3]{3(c)} ($\alpha=1$), with full strength of DMI in all three plots ($\beta =1$). One can clearly see that by the DMI the spins get fully pinned in-plane (see also Fig.~\hyperref[fig3]{3(f)}), and spiral along the current direction, while the strong Heisenberg term renders the spiral wavelength longer. $S_x$ can be fitted perfectly by a single-period trigonometric function. The net magnetization averages out in all directions when the system size is large enough compared to the spiral wavelength.
Fig.~\hyperref[fig3]{3(d)} shows the spin configurations under a modest magnetic field \(H = 0.135\)T, which is sufficient to tilt the spin out-of-plane, leading to a net magnetization $\overline{S_z}$ out-of-plane, while the in-plane spiral texture is unchanged (c.f. Fig.~\hyperref[fig3]{3(g)}). Fig.~\hyperref[fig3]{3(e)} plots $\overline{S_z}$ as a function of out-of-plane magnetic field. The modest critical field $H_c$ for the saturation of $\overline{S_z}$ reflects the small magnitude of DMI.

The dependencies of the spiral wavelength $\lambda$ and critical field on the hole filling factor $\nu$ and the IX current are examined in Fig.~\hyperref[fig3]{3(h, i)} (see Supplementary~\cite{supplementary} for the charge orders assumed at the fractional fillings). As expected, the spiral wavelength decreases with the increase of IX current represented by $\delta k$. The increase of $H_c$ with $\nu$ is a consequence of the long range nature of the DMI.
At larger $\nu$, more neighbors contribute to the DMI effective in-plane field on each spin, which thus requires a larger critical magnetic field to align along $z$ direction.
We note that the perpendicular magnetic anisotropy plays a similar role as the magnetic field Zeeman term, setting the lower bound of DMI strength for the emergence of in-plane magnetic spiral.
At small $\delta k$ and $\nu$, the vanishing $H_c$ and $\lambda^{-1}$ mean the ground states become out-of-plane ferromagnetic. Increasing the exciton coherence length (hence the effective DMI range along current direction) can also turn the magnetic order into in-plane spiral under these conditions.



\begin{figure}[htpb]
    \centering
    \includegraphics[width=8cm]{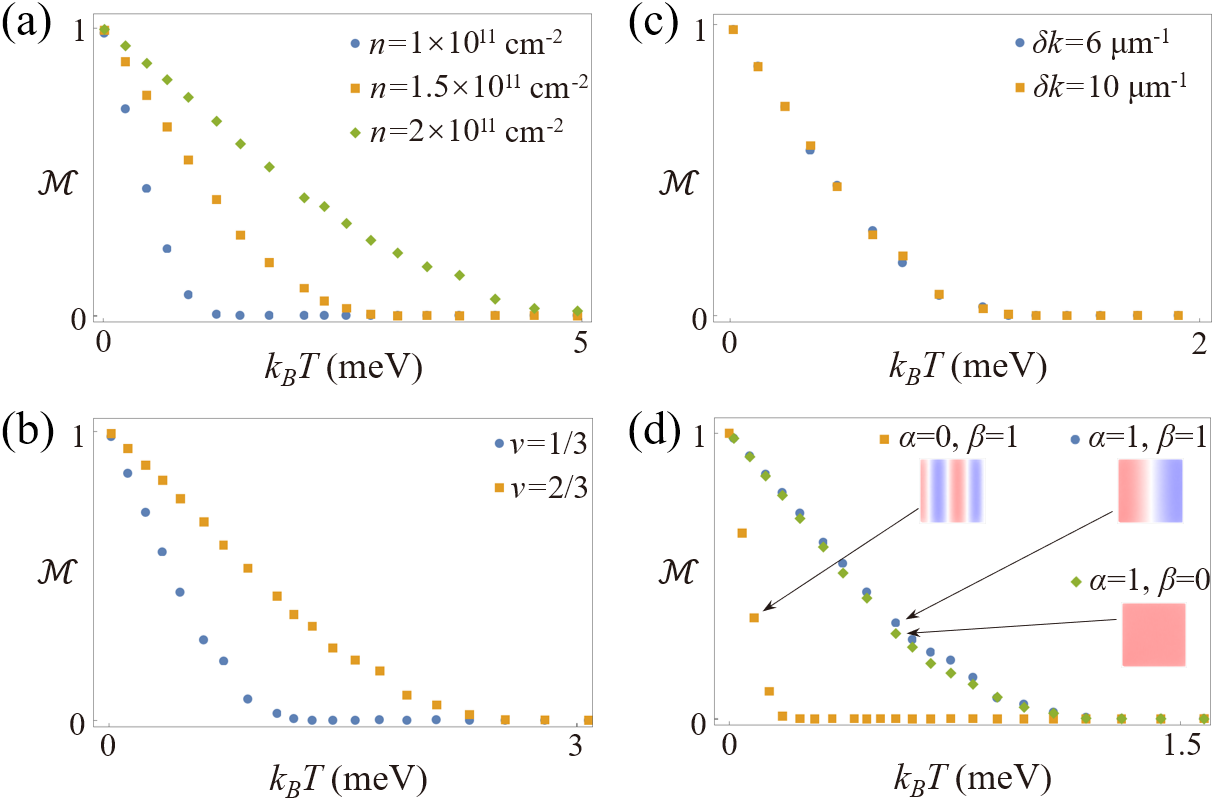}
    \caption{(a-c) Temperature dependence of the magnetic order parameter of moir\'e trapped holes, under different exciton density \(n\) (a), hole filling factor \(\nu\) (b), and IX current denoted by \(\delta k\) (c). The common parameters are \(n=10^{11}\text{cm}^{2}, \delta k =10\mu\)m\(^{-1}\) and \(\nu = 1/3\), unless specified otherwise on the figure panels. (d) Comparison of temperature dependencies when Heisenberg or DMI interaction is switched off. Blue dot: both interactions on ($\alpha=1, \beta=1$). Orange square: Heisenberg switched off ($\alpha=0$). Green diamond: DMI switched off ($\beta=0$). The insets plot spin-spin correlation, showing diminished order parameter at finite temperature whereas the spirals persist.}\label{fig4}
\end{figure}

Finally, we examine the temperature dependence of the magnetic spiral.
The order parameter \(\mathcal{M}\) is defined as the amplitude of the spin-spin correlation which is found in a trigonometric form \(\langle \bm{S}_{i}\cdot\bm{S}_{j}\rangle=\mathcal{M} \cos(\bm{k}\cdot\bm{r}_{ij})\), where the bracket means thermal average.
In Fig.~\hyperref[fig4]{4(a-c)}, the \(\mathcal{M} \)-\(T\) curve is shown under varies IX density, current, and hole filling, which exhibits a critical temperature $k_B T_c$ in the range of few meV.
The increase of $T_c$ with IX density $n$ as shown in Fig.~\hyperref[fig4]{4(a)} is well expected as the latter controls the strength of mediated spin-spin interaction. $T_c$ is also higher at larger filling factor (Fig.~\hyperref[fig4]{4(b)}), which is another reflection of the long interaction range. Noteworthily, $T_c$ has no appreciable change with $\delta k$ (Fig.~\hyperref[fig4]{4(c)}), although the latter tunes the DMI strength dramatically and hence the spiral wavelength. The observation implies that $T_c$ is dominantly determined by the Heisenberg part which is not sensitive to $\delta k$.

This is confirmed by the simulation where we artificially switch off the Heisenberg or DMI part respectively by setting \(\alpha\) or \(\beta\) to zero in Eq.~(\ref{hamil}).
As shown in Fig.~\hyperref[fig4]{4(d)}, when the DMI is turned off, the \(\mathcal{M} \)-\(T\) curve is essentially unchanged, except that the magnetic order changes from the in-plane spiral to the out-of-plane ferromagnetic at \(\beta=0\). In contrast, when Heisenberg is turned off, \(T_c\) becomes small.
Moreover, upon increasing temperature close to $T_c$, \(\mathcal{M} \) diminishes but the magnetic spiral structure persists. These comparisons suggest that the long-wavelength magnetic spiral is a unique ferromagnetic texture jointly formed by the strong but shorter-range Heisenberg and the ultralong-range DMI, in a mesoscopic superlattice. As the interactions are fully controlled by the optically and electrically injectable IX and its spin current, the findings point to an exciting possibility to dynamically establish and manipulate magnetization textures in Mott-Wigner crystal states in moir\'e.

\begin{acknowledgments}
    We thank Xu Zhang, Xi Wang, Jiayi Zhu and Xiaodong Xu for stimulating and helpful discussions. The work is support by the Research Grant Council of Hong Kong SAR (AoE/P-701/20, HKU SRFS2122-7S05), the Croucher Foundation, and the National Natural Science Foundation of China (No. 12074195). W.Y. acknowledges support by Tencent Foundation. Y.W. also wishes to thank the host of Kavli Institute for Theoretical Sciences at UCAS.
\end{acknowledgments}

\bibliographystyle{apsrev4-1}

%

\end{document}